\documentclass[preprint,nofootinbib,letterpaper,aps,amsmath,amsfonts,12pt]{revtex4}
\begin{document}

\setlength\arraycolsep{2pt} 
\def\Pcm#1{{\mathcal{#1}}}
\def\nn{\nonumber}
\def\er#1{eqn.\eqref{#1}}
\def\fgref#1{fig.\ref{#1}}
\newcommand{\del}{\partial}
\newcommand{\re}{\Re \textup{e} \ }
\newcommand{\im}{\Im \textup{m} \ }
\newcommand{\td}{\textup{d}}
\newcommand{\tr}{\textup{tr}}
 
\title{Four-Impurity Operators and String Field Theory Vertex in the BMN Correspondence}
\author{P.~Matlock}
\email{pwm@sfu.ca}
\affiliation{Department of Physics, Simon Fraser University, Canada}
\author{K.~S.~Viswanathan}
\email{kviswana@sfu.ca}
\affiliation{Department of Physics, Simon Fraser University, Canada}

\begin{abstract}

In the context of the Penrose/BMN limit of the AdS/CFT correspondence,
we consider four-impurity BMN operators in Yang-Mills theory, and
demonstrate explicitly their correspondence to four-oscillator states
in string theory. Using the dilatation operator on the gauge-theory
side of the correspondence, we calculate matrix elements between
four-impurity states. Since conformal dimensions of gauge-theory
operators correspond to light-cone energies of string states, these
matrix elements may be compared with the string-theory light-cone
Hamiltonian matrix elements calculated in the plane-wave background
using the string field theory vertex. We find that the two
calculations agree, extending the cases of two- and three-impurity
operators considered in the literature using BMN gauge-theory quantum
mechanics. The results are also in agreement with calculations in the
literature based on perturbative gauge-theory methods.

\end{abstract}

\maketitle


\section{Introduction}

The conjectured AdS/CFT correspondence\cite{9711200,9802150,9905111} between 
Superstring theory and Super Yang-Mills theory has withstood many tests during the 
years since it was put forward. Since it has not been possible to verify AdS/CFT 
directly, different special cases have been considered and elucidated. One such
limit is the well-known `BMN limit,' proposed by Berenstein, Maldacena and 
Nastase \cite{0202021}. On the string side of the 
correspondence, a Penrose limit of AdS space is taken, producing a plane-wave 
background\cite{0110242}. On the Yang-Mills side this corresponds to a specific sector of
the theory in which both $N$ and $J$, the $R$-charge, are taken to infinity, 
holding $N/J^2$ fixed. This is known as the BMN double-scaling 
limit\cite{0205033,0205089,0208178,0209002,0212269,0211032,0307032}. A nice review 
has been given in \cite{0310119}.
To leading order in $g_{\textup{YM}}^2N/J^2$ the light-cone gauge energy 
of a string state is given by
\begin{equation}
E_{\textup{LC}}/\mu = \Delta-J
\end{equation}
where $\mu$ is the mass parameter of the AdS space, $\Delta$ is the
scaling dimension of the gauge-theory operator corresponding to the string state,
and J is its $R$-charge. Writing the exact correspondence in terms of operators, 
the light-cone string Hamiltonian is given by
\begin{equation}
\Pcm{H}_{\textup{LC}}/\mu = D-J
\end{equation}
where $D$ is the dilatation operator in the Yang-Mills theory. 
In the present paper, we consider the bosonic sector of this correspondence.

On the Yang-Mills side, one may write down a perturbative expansion in the
effective t~'Hooft coupling $\lambda'=g_{\textup{YM}}^2N/J^2$ and in
the genus-counting parameter $g_2=J^2/N$. The gauge theory sports an
$SO(6)$ R-symmetry group with fields $\phi_m$ transforming in the
vector representation. One picks two of these, say $\phi_5$ and
$\phi_6$, and defines $Z=\phi_5 + i \phi_6$ and $\bar{Z}=\phi_5 - i\phi_6$, 
which have plus and minus unit $R$-charge respectively.  The
operators in correspondence with oscillator string states in this
limit are the BMN operators, which are products of traces of
powers of $Z$, sprinkled with `impurities' which consist of the other
four $\phi$ fields\cite{0202021,0305016,0307033}. In the BMN limit,
these powers are taken to be large and these operators form a basis in
which one can investigate the dilatation operator $D$ and obtain
information about the light-cone string Hamiltonian. The number of
impurities $h$ in such an operator is identified with the excitation level
of the corresponding string state, and the total number of fields ($Z$
and $\phi$) corresponds to $J+h$, which is the $R$-charge and
engineering dimension of the operator. Anomalous dimensions of these
operators correspond to light-cone string
energies.\cite{0202021,0205066,0306139,0206079} Finding anomalous
conformal dimensions amounts to diagonalising the dilatation operator,
and this is made non-trivial by operator mixing \cite{0209002} when
non-planar and one-loop contributions are considered.  
Such investigations may also be carried out by
considering two-point functions \cite{0205089,0302064,0210153,0306234}.

In the present paper we investigate four-impurity BMN operators,
using the dilatation operator to construct the string Hamiltonian for
level-four states.  This method, called ``BMN Quantum Mechanics'' has been used in the
two-impurity\cite{0303220} and three-impurity\cite{0402211} cases.
Since there are only four distinct impurity fields, more impurities
cannot be considered without taking into account the
combinatoric effects of repeated impurities; thus we complete the
analysis of possible distinct-impurity states by considering
the maximal four-impurity case.
Operators with arbitrary numbers of impurities have been considered
using Yang-Mills perturbation theory\cite{0306234} in which matrix elements
can be extracted from three-point functions. This method
has been used for various combinations of scalar and vector impurities\cite{0301250,0306234}
and for the case of two fermion impurities\cite{0403188}.
Our results using BMN Quantum Mechanics will be found to agree with
these perturbative Yang-Mills computations. 
Using the perturbative approach it was possible for the authors of \cite{0301250} and \cite{0306234}
to consider an arbitrary number of impurities; from our experience
with the four-impurity calculation, it seems tedious to consider 
higher-impurity states using the quantum-mechanical method.
In contrast, it should be mentioned
that since in the perturbative approach the three-point function
is used to obtain matrix elements, the quantum-mechanical
approach may lend itself more readily to the consideration of 
multitrace states with more than two traces.

We construct the string Hamiltonian by first diagonalising the
dilatation operator at leading order and planar level, and then using
this basis to write genus-one dilatation operator elements at one-loop
order.  The calculations, although analogous to two- and
three-impurity cases, are very lengthy and tedious in comparison.  The
resulting Hamiltonian, expressed in terms of its matrix elements in
the four-impurity state basis is expected to be in correspondence with
the three-string light-cone interaction vertex calculated in String
Field Theory. We calculate this vertex for level-four string states
and find precise agreement with the gauge-theory calculation; the
string interaction Hamiltonian of three string states with a total of
four excitations is verified to correspond to the matrix dilatation
operator on the gauge-theory side between the corresponding
four-impurity BMN operators.


\section{Four-Impurity BMN Operators}
Using the notation
\begin{equation}
\Pcm{O}^{1,2,\dots,h}_{p_1,p_2,\dots,p_h} \equiv \tr[ \phi_1 Z^{p_1} \phi_2 Z^{p_2} ... \phi_h Z^{p_h} ]
\end{equation}
for a single-trace operator with $h$ impurities\footnote{As mentioned 
in the introduction, for more that four 
impurities, the same $\phi$ fields would have to be repeated.},
$l$-trace four-impurity BMN operator basis elements may be written
\begin{eqnarray}
\label{dbasisfour}
\textup{`4'}&:& \qquad \Pcm{O}_{p_1p_2p_3p_4}^{1234}\prod_{j=1}^l\Pcm{O}_{J_j}
\\
\label{dbasisthreeone}
\textup{`31'}&:& \qquad\Pcm{O}_{p_1p_2p_3}^{123}\Pcm{O}_{p_4}^{4}\prod_{j=1}^l\Pcm{O}_{J_j} 
\\
\label{dbasistwotwo}
\textup{`22'}&:& \qquad\Pcm{O}_{p_1p_2}^{12}\Pcm{O}_{p_3p_4}^{34}\prod_{j=1}^l\Pcm{O}_{J_j}
\\
\label{dbasistwooneone}
\textup{`211'}&:& \qquad\Pcm{O}_{p_1p_2}^{12}\Pcm{O}_{p_3}^{3}\Pcm{O}_{p_4}^{4}\prod_{j=1}^l\Pcm{O}_{J_j}
\\
\label{dbasisoneoneoneone}
\textup{`1111'}&:& \qquad\Pcm{O}_{p_1}^{1}\Pcm{O}_{p_2}^{2}\Pcm{O}_{p_3}^{3}\Pcm{O}_{p_4}^{4}\prod_{j=1}^l\Pcm{O}_{J_j}
\end{eqnarray}
where $\Pcm{O}_p$ simply indicates an operator with no impurities, $\tr[Z^p]$.
We shall use the indicated labels `4', `31', `22', `211' and `1111' as a short-hand 
way of referring to these operators.
It may be noted that the superscripts in the above, in addition to indicating which 
impurities are present, also denote the order of the impurities.
We write $J_0=p_1+p_2+p_3+p_4$ so that $J=\sum_{j=1}^l J_j$ is the total $R$-charge of the operator.
Taking the continuum BMN limit $J\rightarrow\infty$ with $x_i = p_i/J$ and $r_i = J_i/J$, 
string states corresponding to the above will be denoted as
\begin{eqnarray}
\label{xbasis}
\textup{`4'}&:& \qquad\frac{\sqrt{N^{J+4}}}{J}| x_1,x_2,x_3,x_4 \rangle^{1234} \prod_{j=1}^l | r_j \rangle
\\
\textup{`31'}&:& \qquad\frac{\sqrt{N^{J+4}}}{J}| x_1,x_2,x_3 \rangle^{123} | x_4 \rangle^4 \prod_{j=1}^l | r_j \rangle
\\
\textup{`22'}&:& \qquad\frac{\sqrt{N^{J+4}}}{J}| x_1,x_2 \rangle^{12} | x_3,x_4 \rangle^{34} \prod_{j=1}^l | r_j \rangle
\\
\textup{`211'}&:& \qquad\frac{\sqrt{N^{J+4}}}{J}| x_1,x_2 \rangle^{12} |x_3 \rangle^3 | x_4 \rangle^4 \prod_{j=1}^l | r_j \rangle
\\
\textup{`1111'}&:& \qquad\frac{\sqrt{N^{J+4}}}{J}| x_1 \rangle^1 | x_2 \rangle^2 |x_3 \rangle^3 | x_4 \rangle^4 \prod_{j=1}^l | r_j \rangle
\end{eqnarray}
The normalisation of these states may be understood by examining their tree-level planar two-point
functions\cite{0402211}.
\section{String Hamiltonian}
We wish to find the string Hamiltonian, given by
\begin{equation}
H=\lim_{N\rightarrow\infty, N/J^2 \textup{fixed}} (D-J)
,\end{equation} in the basis defined in the previous section.
As explained in \cite{0402211} this is not Hermitian, so that $H$ found in
this way may not be directly interpreted as the string Hamiltonian. To
remedy this, one begins by defining the inner product using the planar
free theory,
\begin{equation}
\langle a | b \rangle \equiv \langle \Pcm{O}_a \Pcm{O}_b \rangle_{\textup{free, planar}}
;\end{equation}
$H$ is not Hermitian with respect to this product, but with respect to the
product defined by the full non-planar free correlator
\begin{equation}
\langle a | b \rangle_{g_2} \equiv \langle \Pcm{O}_a \Pcm{O}_b \rangle_{\textup{free, full}}
 \equiv \langle a | S | b \rangle
,\end{equation}
where $S$ is Hermitian with respect to the original planar product. 
A new basis state may be defined by the non-unitary transformation
\begin{equation}
 |\tilde{a}\rangle \equiv S^{-1/2}|a\rangle
.\end{equation}
Now, a `new' Hamiltonian $\tilde{H}$ may be defined by\cite{0210102,0210153,0206059,0209215}
\begin{equation}
\langle\tilde{a}| \tilde{H} | \tilde{b}\rangle \equiv 
\langle\tilde{a}| H | \tilde{b}\rangle_{g_2}
\end{equation}
so that
\begin{equation}
\tilde{H}=S^{1/2}HS^{-1/2}
\end{equation}
is Hermitian in the original basis and should correspond to the
light-cone string Hamiltonian $\Pcm{H}_{\textup{LC}}$, up to a possible unitary
transformation.  Now, since $\langle a | S | b \rangle = \langle a | b \rangle_{g_2}$,
$S$ is just the full non-planar mixing matrix for basis states. 
Expanding $S$ and $H$ in the genus-counting parameter $g_2$,
\begin{equation}
S=1+g_2\Sigma + \Pcm{O}(g_2^2) \quad,\quad H=H_0+g_2H_1+\Pcm{O}(g_2^2)
,\end{equation}
and we may write matrix elements of the string Hamiltonian;
\begin{equation}
\label{Htildeform}
\langle a | \tilde{H} | b \rangle = 
\langle a | (1+\frac12g_2\Sigma)H(1-\frac12g_2\Sigma)| b \rangle
=\langle a | H_0 | b \rangle 
+ g_2 \langle a | \big( \frac12 [\Sigma,H_0] +H_1 \big)| b \rangle
\end{equation}
Here, $H_0$ is just the planar part of $D_2$, the dilatation operator at one-loop order
while $H_1$ is the genus-one part of $D_2$. The dilatation operator is given 
by \cite{0205033,0208178,0209002,0303060,0307015,0310252,0310232}
\begin{equation}
D=D_0 + \frac{g_{\textup{YM}}^2}{16\pi^2}D_2 + \Pcm{O}(g_{\textup{YM}}^4)
\end{equation}
where
\begin{equation}
D_0=\tr\big( \phi_m \check{\phi}_m \big) \quad , \quad
  D_2 = -:\tr \big( 
 [\phi_m,\phi_n] [\check{\phi}_m,\check{\phi}_n]
+\frac12 [\phi_m,\check{\phi}_n] [\phi_m,\check{\phi}_n]
\big):
.\end{equation}
Here, $m$ and $n$ are $SO(6)$ indices running from $1$ to $6$,
$\check{\phi}=\delta/\delta\phi$ and the normal ordering symbol denotes that the
enclosed $\phi$-derivatives only act on fields outside it. 
For operators containing $Z$ and $\phi_i$ fields 
(but no $\bar{Z}$ fields), the above expression for $D_2$ becomes
\begin{equation}
\label{dill}
  D_2 = -:\tr \big( 
 [\phi_i,\phi_j] [\check{\phi}_i,\check{\phi}_j]
+2 [\phi_i,Z] [\check{\phi}_i,\check{Z}]
+\frac12 [\phi_i,\check{\phi}_j] [\phi_i,\check{\phi}_j]
\big):
\end{equation}
where $i$ and $j$ run only over impurity fields, that is from $1$ to $4$.
In the following we do not consider `boundary' terms in which the impurities 
are neighbours\footnote{Such terms make diagonalisation of $D$ very difficult 
in a discrete basis, but in the continuum BMN limit they become unimportant.},
so that the first term in \er{dill}, which always produces 
such states, may be neglected. Since the four impurities we consider are 
distinct, the third term may also be neglected, since it contains a
double derivative of a single $\phi$ field.\footnote{These would contribute in 
the case of more than four impurities.}

It will be helpful to make note of the following contraction identities.
\begin{eqnarray}
\textup{tr}[1]&=&N, \\
\textup{tr}[A\check{\phi}]\textup{tr}[B\phi]&=&\textup{tr}[AB] \qquad \textup{ (``fusion'')},\\
\textup{tr}[A\check{\phi}B\phi]&=&\textup{tr}[A]\textup{tr}[B] \qquad \textup{ (``fission'')},
\end{eqnarray}
and the complete contractions
\begin{eqnarray}
\textup{tr}[Z^p\bar{Z}^q]&=&\delta_{pq}N^{p+1} + \Pcm{O}(N^{p-1}), \\
\textup{tr}[Z^p]\textup{tr}[\bar{Z}^qZ^r]&=&\delta_{p+r,q}p(r+1)N^{p+r} + \Pcm{O}(N^{p+r-2}), \\
\textup{tr}[Z^p\bar{Z}^q{Z}^r\bar{Z}^s]&=&\delta_{p+r,q+s}N^{p+r+1}(\textup{min}(p,q,r,s)+1)
 + \Pcm{O}(N^{p+r-1}).
\end{eqnarray}

Operating with $H_0$ on the discrete basis \eqref{dbasisfour}-\eqref{dbasisoneoneoneone}
we find
\begin{eqnarray}
H_0\Pcm{O}_{p_ap_bp_cp_d}^{abcd}\prod_j\Pcm{O}_{J_j} &=& -\frac{g^2_{\textup{YM}}N}{8\pi^2}\big(
	       -2\Pcm{O}^{abcd}_{p_a,p_b,p_c,p_d}
	       +\Pcm{O}^{abcd}_{p_a-1,p_b,p_c,p_d+1}
	       +\Pcm{O}^{abcd}_{p_a+1,p_b,p_c,p_d-1}
          \big)\prod_j\Pcm{O}_{J_j} \nn\\
         & +& (\text{three other cyclic permutations of }abcd) \\
H_0\Pcm{O}_{p_ap_bp_c}^{abc}\Pcm{O}^d_{p_d}\prod_j\Pcm{O}_{J_j} &=& -\frac{g^2_{\textup{YM}}N}{8\pi^2}\big(
	       -2\Pcm{O}^{abc}_{p_a,p_b,p_c}
	       +\Pcm{O}^{abc}_{p_a-1,p_b,p_c+1}
	       +\Pcm{O}^{abc}_{p_a+1,p_b,p_c-1}
          \big) \Pcm{O}^d_{p_d} \prod_j\Pcm{O}_{J_j} \nn\\
         & +& (\text{two other cyclic permutations of }abc) \\
H_0\Pcm{O}_{p_ap_b}^{ab}\Pcm{O}^{cd}_{p_cp_d}\prod_j\Pcm{O}_{J_j} &=& -\frac{g^2_{\textup{YM}}N}{2\pi^2}\big(
	       -2\Pcm{O}^{ab}_{p_a,p_b}
	       +\Pcm{O}^{ab}_{p_a-1,p_b+1}
	       +\Pcm{O}^{ab}_{p_a+1,p_b-1}
          \big) \Pcm{O}^{cd}_{p_cp_d} \prod_j\Pcm{O}_{J_j} \nn\\
   &+& ( abcd \rightarrow cdab ) \\
H_0\Pcm{O}_{p_ap_b}^{ab}\Pcm{O}_{p_c}^{c}\Pcm{O}_{p_d}^{d}\prod_j\Pcm{O}_{J_j} &=& -\frac{g^2_{\textup{YM}}N}{4\pi^2}
             \big(
	       -2\Pcm{O}^{ab}_{p_a,p_b}
	       +\Pcm{O}^{ab}_{p_a-1,p_b+1}
	       +\Pcm{O}^{ab}_{p_a+1,p_b-1}
             \big)\Pcm{O}_{p_c}^{c}\Pcm{O}_{p_d}^{d} \prod_j\Pcm{O}_{J_j} \\
H_0\Pcm{O}_{p_a}^{a}\Pcm{O}_{p_b}^{b}\Pcm{O}_{p_c}^{c}\Pcm{O}_{p_d}^{d}\prod_j\Pcm{O}_{J_j} &=& 0
.\end{eqnarray}
In the continuum limit these become ($x_{123}$ denotes $x_1+x_2+x_3$, etc.)
\begin{eqnarray}
&&H_0 | x_1,x_2,x_3,r_0-x_{123} \rangle^{1234}  \nn\\
&&\quad=-\frac{\lambda'}{8\pi^2}\big(
\partial_{x_1}^2+(\partial_{x_2}-\partial_{x_1})^2+(\partial_{x_3}-\partial_{x_2})^2+\partial_{x_3}^2
 \big)
| x_1,x_2,x_3,r_0-x_{123} \rangle^{1234}  \\
&&H_0 | x_1,x_2,x_3\rangle^{123} |r_0-x_{123} \rangle^{4}  \nn\\
&&\quad=-\frac{\lambda'}{8\pi^2}\big(
\partial_{x_1}^2+\partial_{x_2}^2+(\partial_{x_2}-\partial_{x_1})^2
 \big)
| x_1,x_2,x_3\rangle^{123} |r_0-x_{123} \rangle^{4}  \\
&&H_0 | x_1,x_2 \rangle^{12} |x_3,r_0-x_{123} \rangle^{34} \nn\\
&&\quad=-\frac{\lambda'}{8\pi^2}\big(
(\partial_{x_1}-\partial_{x_2})^2+\partial_{x_3}^2
 \big)
| x_1,x_2 \rangle^{12} |x_3,r_0-x_{123} \rangle^{34} \\
&&H_0 | x_1,x_2 \rangle^{12} |x_3 \rangle^3 |r_0-x_{123} \rangle^4 
=-\frac{\lambda'}{4\pi^2}
\partial_{x_1}^2
| x_1,x_2 \rangle^{12} |x_3 \rangle^3 |r_0-x_{123} \rangle^4 
\end{eqnarray}
In the above expressions, we have suppressed the factor $\prod_{j=1}^l | r_j \rangle$
since it is unaffected by $H_0$.
The above eigenvalue equations are solved by defining the following 
momentum-basis states\cite{0205089,0302064}. 
\begin{eqnarray}
\label{mbasisfour}
&&|n_1,n_2,n_3;r_0\rangle  \equiv \frac1{\sqrt{r_0^3}} \nn\\
&&\quad\times \sum_{(abc)} \int_{x_{123}<r_0} \td^3x
e^{\frac{2\pi i}{r_0}(n_ax_1+n_bx_{12}+n_cx_{123})}
\Pcm{O}^{1,a+1,b+1,c+1}(x_1,x_2,x_3,r_0-x_{123}) ,
\end{eqnarray}
\begin{eqnarray}
\label{mbasisthreeone}
&&|n_1,n_2;r_0-s\rangle^{123} |s\rangle^4 \equiv \frac1{r_0-s} \nn\\ 
&&\quad\times \sum_{(ab)} \int_{x_{12}<r_0-s} \td^2x
e^{\frac{2\pi i}{r_0-s}(n_ax_1+n_bx_{12})}
\Pcm{O}^{1,a+1,b+1}(x_1,x_2,r_0-s-x_{12}) \Pcm{O}^{4}(s) ,
\end{eqnarray}
\begin{eqnarray}
\label{mbasistwotwo}
&&|n_1;r_0-s\rangle^{12} |n_2;s\rangle^{34} \equiv \frac1{\sqrt{(r_0-s)s}} \nn\\
&&\quad\times \int_0^{r_0-s} \td x_1 \int_0^{s} \td x_2
e^{2\pi i(\frac{n_1}{r_0-s}x_1+\frac{n_b}{s}x_{2})}
\Pcm{O}^{12}(x_1,r_0-s-x_1)\Pcm{O}^{34}(x_2,s-x_2) ,
\end{eqnarray}
\begin{eqnarray}
\label{mbasistwooneone}
&&|n;r_0-s-t\rangle^{12} |s\rangle^3 |t\rangle^4 \equiv \frac1{\sqrt{r_0-s-t}} \nn\\
&&\quad\times \int_0^{r_0-s-t} \td x
e^{\frac{2\pi i}{r_0-s-t}nx}
\Pcm{O}^{12}(x,r_0-s-t-x)\Pcm{O}^{3}(s)\Pcm{O}^{4}(t)
,\end{eqnarray}
while zero-impurity states are normalised as $|r\rangle \equiv \frac1{\sqrt{r}}\Pcm{O}(r)$.
Here, $\Pcm{O}(x)$ simply denotes the continuum version of the discrete 
operator basis \eqref{dbasisfour}-\eqref{dbasisoneoneoneone}.
$\sum_{(abc)}$ denotes a summation in which $abc$ takes on each of the
six permutations of $123$, and similarly for $\sum_{(ab)}$.
The superscripts now indicate only which impurities are contained 
in each trace.
These above momentum states have the energy eigenvalues
\begin{eqnarray}
\label{Hnoughtff}
E_\textup{`4'} &=& \frac{\lambda'}{2} \frac{n_{123}^2 +n_1^2+n_2^2+n_3^2}{r_0^2}, \\
E_\textup{`31'} &=& \frac{\lambda'}{2} \frac{n_{12}^2 +n_1^2+n_2^2 }{(r_0-s)^2}, \\
E_\textup{`22'} &=& \frac{\lambda'}{2} \bigg( \frac{n_1^2}{(r_0-s)^2} + \frac{n_2^2}{s^2} \bigg), \\
E_\textup{`211'} &=& {\lambda'} \frac{n^2}{(r_0-s-t)^2}, \\
E_\textup{`1111'} &=& 0.
\end{eqnarray}
Again, these do not depend on whether or not the states $\prod | r_j \rangle$ are present; $H_0$
will not operate on such factors.
The momentum states \eqref{mbasisfour}-\eqref{mbasistwooneone} are orthonormal, so that
\begin{equation}
\prod_{j=1}^{l}\langle r_j|m_1,m_2,m_3;r_0|n_1,n_2,n_3;s_0\rangle\prod_{j=1}^{l'}| s_j\rangle
=\delta_{m_1,n_1}\delta_{m_2,n_2}\delta_{m_3,n_3}\delta_{l,l'}\delta(r_0-s_0)
\sum_{\pi\in S_l}\prod_{k=1}^l \delta(r_{\pi(k)}-s_k)
\end{equation}
and similarly for the remaining states \eqref{mbasisthreeone}-\eqref{mbasistwooneone}.

\setlength\arraycolsep{0pt}
We must now calculate explicitly $H_1$ and $\Sigma$ in the above $H_0$-eigenstate basis.
For clarity, we follow the convention of writing $H_1=H_++H_-$, where 
$H_+$ increases and $H_-$ decreases the number of traces when 
acting on a basis state.
The procedure is several times more tedious and lengthy than in the 
three-impurity case \cite{0402211}, and we relegate the explicit computation
of these quantities to Appendices I and I\!I.

The calculation of $H_1$ in Appendix I leads us to the following observations.
Upon writing out the matrix elements of $H_+$ and $H_-$, we
see that for matrix elements not involving a `4'-state,
the calculation effectively reduces to the case of fewer impurities.
The matrix elements between our `31', `211', and `1111' states therefore
need not be considered further, since they correspond to the three-impurity 
case already studied in \cite{0402211} or to the two-impurity calculation of \cite{0303220}.
Moreover, we note that the `22-22' matrix element of $H_1$ simply involves
two copies of the `2-2' element from the two-impurity case studied 
in \cite{0303220}.

We point out that
in the following, the notations $D_a$ and $D_a^b$ are used differently 
in each case. 
We continue to omit the product $\prod_j|r_j\rangle$ of 
zero-impurity states which we omitted in the above, with
the understanding that this product may appear in each of the following
states without affecting the calculation.
Referring again to Appendix I, the `4-4' elements of $H_+$ and $H_-$ are given by
\begin{eqnarray}
\label{hpff}
 \langle s | &&\langle m_1,m_2,m_3;r_0 -s| H_+ | n_1,n_2,n_3;r_0\rangle \nn\\
   && = \frac{-\lambda'}{2\pi^4} \frac{1}{\sqrt{(r_0-s)^3sr_0^3}}
       \sin\big(\frac{\pi s n_1}{r_0}\big)
       \sin\big(\frac{\pi s n_2}{r_0}\big)
       \sin\big(\frac{\pi s n_3}{r_0}\big)
       \sin\big(\frac{\pi s n_{123}}{r_0}\big) \nn\\
&&\times \frac{m_1D_1+m_2D_2+m_3D_3+m_{123}D_{123}}
	      {(r_0-s)D_1D_2D_3D_{123}} \nn\\
\textup{where } && D_a = \frac{n_a}{r_0} - \frac{m_a}{r_0-s}
\end{eqnarray}
and
\begin{eqnarray}
\label{hmff}
 &&\langle m_1,m_2,m_3;r_0 | H_- | n_1,n_2,n_3;r_0-s\rangle |s\rangle \nn\\
    &&= \frac{\lambda'}{2\pi^4} \frac1{\sqrt{(r_0-s)^3sr_0^3}}
      \sin\big(\frac{\pi s m_1}{r_0}\big)
      \sin\big(\frac{\pi s m_2}{r_0}\big)
      \sin\big(\frac{\pi s m_3}{r_0}\big)
      \sin\big(\frac{\pi s m_{123}}{r_0}\big) \nn\\
&&\times 
      \left[ 
	\frac{m_{123}}{r_0D_{123}}\left(\frac1{D_2D_3}+\frac1{D_1D_3}+\frac1{D_1D_2}\right)
	+\frac1{r_0D_{123}}\left(\frac{m_1}{D_2D_3}+\frac{m_2}{D_1D_3}+\frac{m_3}{D_1D_2}\right)
      \right]
\nn\\
\textup{where }&& D_a = \frac{n_a}{r_0-s} - \frac{m_a}{r_0}
.\end{eqnarray}
For the `4-31' elements of $H_+$ and $H_-$ we find
\begin{eqnarray}
\label{hpfto}
 &&{}^{123}\langle m_1,m_2;r_0 - s| {}^4\langle s | H_+ | n_1,n_2,n_3;r_0\rangle \nn\\
   &&= \frac{-\lambda'}{2\pi^4} \frac{1}{\sqrt{(r_0-s)^2r_0^3}}
       \sin\big(\frac{\pi s n_1}{r_0}\big)
       \sin\big(\frac{\pi s n_2}{r_0}\big)
       \sin\big(\frac{\pi s n_3}{r_0}\big)
       \sin\big(\frac{\pi s n_{123}}{r_0}\big) \nn\\
&&\times 
        \left[
	\frac{m_1}{r_0-s}\left(\frac1{D_2^{-2}D_1^{-1}}+\frac1{D_{123}^{-12}D_2^{-2}}\right)
	+ \frac{m_2}{r_0-s}\left(\frac1{D_2^{-2}D_1^{-1}}+\frac1{D_{123}^{-12}D_1^{-1}}\right)
	\right] \nn\\
\textup{where }&& D_a^b = \frac{n_a}{r_0} + \frac{m_b}{r_0-s}
\end{eqnarray}
and
\begin{eqnarray}
\label{hmfto}
 &&\langle m_1,m_2,m_3;r_0 | H_- | n_1,n_2;r_0-s\rangle^{123}|s\rangle^4\nn\\
   &&=  \frac{\lambda'}{2\pi^4} \frac{1}{\sqrt{(r_0-s)^2r_0^3}} 
       \sin\big(\frac{\pi s m_1}{r_0}\big)
       \sin\big(\frac{\pi s m_2}{r_0}\big)
       \sin\big(\frac{\pi s m_3}{r_0}\big)
       \sin\big(\frac{\pi s m_{123}}{r_0}\big) \nn\\
&&\times 
        \left[
	  \left(\frac1{D_{12}^{-123}}-\frac1{D_{12}^{-12}}-\frac{m_{123}}{m_3D_{12}^{-12}}\right)
	  \left(\frac1{D_1^{-1}}+\frac1{D_2^{-2}}\right)
	  -\frac1{m_3D_{12}^{-123}}\left(\frac{m_1}{D_2^{-2}}+\frac{m_2}{D_1^{-1}}\right)
	\right] \nn\\
\textup{where } &&D_a^b = \frac{n_a}{r_0-s} + \frac{m_b}{r_0}
.\end{eqnarray}
Finally, the `4-22' elements are
\begin{eqnarray}
\label{hpftt}
 &&{}^{12}\langle m_1;r_0 - s| {}^{34}\langle m_2;s | H_+ | n_1,n_2,n_3;r_0\rangle \nn\\
   &&=  \frac{-\lambda'}{2\pi^4} \frac{1}{\sqrt{(r_0-s)sr_0^3}}
       \sin\big(\frac{\pi s n_1}{r_0}\big)
       \sin\big(\frac{\pi s n_2}{r_0}\big)
       \sin\big(\frac{\pi s n_3}{r_0}\big)
       \sin\big(\frac{\pi s n_{123}}{r_0}\big) \nn\\
&&\times 
        \left[
	\frac{m_1}{r_0-s}\frac1{D_2^2D_3^{-2}}\left(\frac1{D_1^{-1}}+\frac1{D_{123}^{-1}}\right)
	+\frac{m_2}{s}\frac1{D_{123}^{-1}D_1^{-1}}\left(\frac1{D_2^{2}}-\frac1{D_{3}^{-2}}\right)
	\right] \nn\\
\textup{where }&& D_a^1 = \frac{n_a}{r_0} + \frac{m_1}{r_0-s} \textup{ and }
                  D_a^2 = \frac{n_a}{r_0} + \frac{m_2}{s}
\end{eqnarray}
and
\begin{eqnarray}
\label{hmftt}
 &&\langle m_1,m_2,m_3;r_0 | H_- | n_1;r_0-s\rangle^{12}|n_2;s\rangle^{34}\nn\\
   &&= \frac{\lambda'}{2\pi^4} \frac{1}{\sqrt{(r_0-s)sr_0^3}}
       \sin\big(\frac{\pi s m_1}{r_0}\big)
       \sin\big(\frac{\pi s m_2}{r_0}\big)
       \sin\big(\frac{\pi s m_3}{r_0}\big)
       \sin\big(\frac{\pi s m_{123}}{r_0}\big) \nn\\
&&\times 
        \left[
	  \frac{m_1-m_2}{m_{23}}\frac1{D_{1}^{-123}D_2^{-3}}
	  +\frac{m_2+m_{23}}{m_{23}}\frac1{D_{1}^{-1}D_2^{-3}}
	  -\frac{m_{123}+m_3}{m_{23}}\frac1{D_{1}^{-1}D_2^{2}}
	  +\frac{m_1-m_3}{m_{23}}\frac1{D_{1}^{123}D_2^{2}}
	\right] \nn\\
\textup{where }&& D_1^b = \frac{n_1}{r_0-s} + \frac{m_b}{r_0} \textup{ and }
                  D_2^b = \frac{n_2}{s} + \frac{m_b}{r_0}
.\end{eqnarray}

Next the matrix elements of $\Sigma$, from Appendix I\!I, are 
given by the following. The `4-4' component of $\Sigma$ is
\begin{eqnarray}
\label{sigff}
 \langle s |
 &&\langle m_1,m_2,m_3;r_0 - s| \Sigma | n_1,n_2,n_3;r_0\rangle \nn\\
    &&= \frac{1}{\pi^4\sqrt{(r_0-s)^3sr_0^3}}
      \sin\big(\frac{\pi s n_1}{r_0}\big)
      \sin\big(\frac{\pi s n_2}{r_0}\big)
      \sin\big(\frac{\pi s n_3}{r_0}\big)
      \sin\big(\frac{\pi s n_{123}}{r_0}\big) \nn\\
&&\times \frac{1}{D_1D_2D_3D_{123}} \nn\\
\textup{where } && D_a = \frac{n_a}{r_0} - \frac{m_a}{r_0-s}
,\end{eqnarray}
while the `4-31' element is found to be
\begin{eqnarray}
\label{sigfto}
 &&{}^{123}\langle m_1,m_2;r_0 - s|{}^4\langle s | \Sigma | n_1,n_2,n_3;r_0\rangle \nn\\
    &&= \frac{1}{\pi^4\sqrt{(r_0-s)^2r_0^3}}
      \sin\big(\frac{\pi s n_1}{r_0}\big)
      \sin\big(\frac{\pi s n_2}{r_0}\big)
      \sin\big(\frac{\pi s n_3}{r_0}\big)
      \sin\big(\frac{\pi s n_{123}}{r_0}\big) \nn\\
&&\times \frac{1}{D_{123}^{-12}D_3D_1^{-1}D_{2}^{-2}} \nn\\
\textup{where } && D_a^b = \frac{n_a}{r_0} + \frac{m_b}{r_0-s}
,\end{eqnarray}
and the `4-22' component is
\begin{eqnarray}
\label{sigftt}
 &&{}^{12}\langle m_1;r_0 - s|{}^{34}\langle m_2; s | \Sigma | n_1,n_2,n_3;r_0\rangle \nn\\
    &&= \frac{1}{\pi^4\sqrt{(r_0-s)sr_0^3}}
      \sin\big(\frac{\pi s n_1}{r_0}\big)
      \sin\big(\frac{\pi s n_2}{r_0}\big)
      \sin\big(\frac{\pi s n_3}{r_0}\big)
      \sin\big(\frac{\pi s n_{123}}{r_0}\big) \nn\\
&&\times \frac{1}{D_{123}^{-1}D_1^{-1}D_3^{-2}D_{2}^{2}} \nn\\
\textup{where }&& D_a^1 = \frac{n_a}{r_0} + \frac{m_1}{r_0-s} \textup{ and }
                  D_a^2 = \frac{n_a}{r_0} + \frac{m_2}{s}
.\end{eqnarray}
\setlength\arraycolsep{2pt}

In the above formulae, other impurity orderings may be accommodated by
considering appropriate permutations of the momenta. These permutations
are found using the definitions of the momentum states \eqref{mbasisfour}-\eqref{mbasistwooneone}.
Let $|\rangle^{abcd}$ be any single- or multi-trace state containing the four 
impurities $abcd$. We wish to find the matrix element of some operator $M$ 
between this state and the `4'-state,
\begin{equation}
{}^{abcd}\langle|M|n_1,n_2,n_3;r_0\rangle
,\end{equation}
given that we already have this quantity for the case $abcd=1234$;
\begin{equation}
M_{n_1,n_2,n_3}\equiv{}^{1234}\langle|M|n_1,n_2,n_3;r_0\rangle
.\end{equation}
The results are, for $\pi\in S_3$ and $abc=\pi(234)$,
\begin{eqnarray}
{}^{1abc}\langle|M|n_1,n_2,n_3;r_0\rangle &=& M_{\pi(n_1,n_2,n_3)} \\
{}^{a1bc}\langle|M|n_1,n_2,n_3;r_0\rangle &=& M_{\pi(-n_{123},n_2,n_3)} \\
{}^{ab1c}\langle|M|n_1,n_2,n_3;r_0\rangle &=& M_{\pi(-n_{123},n_1,n_3)} \\
{}^{abc1}\langle|M|n_1,n_2,n_3;r_0\rangle &=& M_{\pi(-n_{123},n_1,n_2)}
.\end{eqnarray}

Now we are in a position to calculate the string 
Hamiltonian $\tilde{H}$, assembling $H_0$, $H_1$ and $\Sigma$
using \er{Htildeform}.
The `4-4' element of the genus-one correction to $\tilde{H}$ is given by
\begin{equation}
\tilde{H}_\textup{`4-4'}\big|_{g_2}=g_2
\langle s | \langle m_1,m_2,m_3;r_0-s| \big( \frac12 [\Sigma,H_0]+H_1 \big) | n_1,n_2,n_3 ; r_0 \rangle
.\end{equation}
Which may be calculated either using $\Sigma$ and $H_+$ from \er{sigff} and \er{hpff},
or $\Sigma$ and $H_-$ from the conjugate of \er{sigff} and \er{hmff}. It may 
easily be verified that these give the same result, showing that $\tilde{H}$ is Hermitian 
as it should be, and serving as a check on the calculations.
The result is
\begin{eqnarray}
\label{HTff}
\tilde{H}_\textup{`4-4'}\big|_{g_2}
&=&\frac{\lambda'g_2}{4\pi^4}\frac1{\sqrt{(r_0-s)^3sr_0^3}}\frac{(D_1)^2+(D_2)^2+(D_3)^2+(D_{123})^2}{D_1D_2D_3D_{123}} \nn\\
&&\times \sin(\frac{\pi s n_1}{r_0})
\sin(\frac{\pi s n_2}{r_0})
\sin(\frac{\pi s n_3}{r_0})
\sin(\frac{\pi s n_{123}}{r_0}) \nn\\
\textup{where } D_a &=& \frac{n_a}{r_0}-\frac{m_a}{r_0-s}
.\end{eqnarray}
Of course, the genus-zero `4-4' component is simply given by $H_0$ in \er{Hnoughtff}.
The `4-31' element is similarly obtained using $\Sigma$ from \er{sigfto} 
and $H_\pm$ from \er{hpfto} or \er{hmfto} to calculate
\begin{eqnarray}
\label{HTfto}
\tilde{H}_\textup{`4-31'} &=&g_2
{}^{123}\langle m_1,m_2;r_0-s|{}^4\langle s |\big(\frac12 [\Sigma,H_0]+H_1\big)
                       | n_1,n_2,n_3 ; r_0 \rangle \\
&=& \frac{\lambda'g_2}{4\pi^4}\frac1{\sqrt{(r_0-s)^2r_0^3}}
\frac{(D_{123}^{-12})^2+(D_{1}^{-1})^2+(D_{2}^{-2})^2+(D_3)^2}
 {D_{123}^{-12}D_{1}^{-1}D_{2}^{-2}D_3} \nn\\
&&\times \sin(\frac{\pi s n_1}{r_0})
\sin(\frac{\pi s n_2}{r_0})
\sin(\frac{\pi s n_3}{r_0})
\sin(\frac{\pi s n_{123}}{r_0}) \nn\\
\textup{where } && D_a^b = \frac{n_a}{r_0} + \frac{m_b}{r_0-s}
,\end{eqnarray}
and the `4-22' element by using \er{sigftt} and \er{hpftt} or \er{hmftt}, giving
\begin{eqnarray}
\label{HTftt}
\tilde{H}_\textup{`4-22'} &=&g_2
{}^{12}\langle m_1;r_0-s|{}^{34}\langle m_2;s | \big( \frac12 [\Sigma,H_0]+H_1\big)
         | n_1,n_2,n_3 ; r_0 \rangle \\
&=& \frac{\lambda'g_2}{4\pi^4}\frac1{\sqrt{(r_0-s)sr_0^3}}
\frac{(D_{123}^{-1})^2+(D_{2}^{2})^2+(D_{3}^{-2})^2+(D_1^{-1})^2}
{D_{123}^{-1}D_{2}^{2}D_{3}^{-2}D_1^{-1}} \nn\\
&&\times \sin(\frac{\pi s n_1}{r_0})
\sin(\frac{\pi s n_2}{r_0})
\sin(\frac{\pi s n_3}{r_0})
\sin(\frac{\pi s n_{123}}{r_0}) \nn\\
\textup{where }&& D_a^1 = \frac{n_a}{r_0} + \frac{m_1}{r_0-s} \textup{ and }
                  D_a^2 = \frac{n_a}{r_0} + \frac{m_2}{s}
.\end{eqnarray}

With the order-$g_2$ string Hamiltonian now in hand, 
we turn to the computation of the string field theory 
vertex with which it is expected to correspond.

\section{Comparison with String-Field Vertex}

The number of impurities of a BMN operator on the SYM side of the
correspondence is identified with the number of oscillator excitations
of the corresponding state on the string side. Light-cone String Field
Theory in the plane-wave background has been developed 
in \cite{0204146,0208209,0210246,0206073,0208179}, and the Neumann coefficients
necessary for computations have been found in \cite{0211198}
with further results in \cite{0311231,0402185}.

We shall consider a three-string interaction, with a total
of eight oscillator excitations distributed among the three strings.
A multi-string state with $2k$ excitations is given by
\begin{equation}
|A\rangle=\prod_{j=1}^{2k} {\alpha_{(r_j)m_j}^{I_j \dagger}}|0\rangle
\end{equation}
where $r_j$ are the string numbers, $I_j$ label the transverse $AdS$ directions
(i.e. the impurity coordinates), and $m_j$ are the oscillator numbers.
For our purposes, we set $k=4$.

In the case of our `4-4' interaction, we consider the three-string state
\begin{equation}
|\textup{4,4}\rangle=
\alpha_{(1)n_1}^{1\dagger}
\alpha_{(1)n_2}^{2\dagger}
\alpha_{(1)n_3}^{3\dagger}
\alpha_{(1)-n_{123}}^{4\dagger}
\alpha_{(3)m_1}^{1\dagger}
\alpha_{(3)m_2}^{2\dagger}
\alpha_{(3)m_3}^{3\dagger}
\alpha_{(3)-m_{123}}^{4\dagger}
|0\rangle_{(1)}\otimes|0\rangle_{(2)}\otimes|0\rangle_{(3)}
,\end{equation}
where excitations are absent for string number two; 
it corresponds to a zero-impurity state.
In \cite{0307033} it is shown how to calculate the interaction vertex
between the strings in this state. We find
\begin{equation}
\label{SFTff}
\langle\textup{4,4}|H_3\rangle=\frac{\alpha_{(1)}\alpha_{(2)}\alpha_{(3)}}2 \sum_{j=1}^{4}\Pcm{N}_j
\end{equation}
where
\begin{eqnarray}
\Pcm{N}_1&=&\left( \frac{\omega_{(1)n_1}}{\mu\alpha_{(1)}} 
                 + \frac{\omega_{(3)m_1}}{\mu\alpha_{(3)}} \right)
		 \tilde{N}^{(1,3)}_{-n_1,m_1}
		 \tilde{N}^{(1,3)}_{n_2,m_2}
		 \tilde{N}^{(1,3)}_{n_3,m_3}
		 \tilde{N}^{(1,3)}_{-n_{123},-m_{123}} ,\\
\Pcm{N}_2&=&\left( \frac{\omega_{(1)n_2}}{\mu\alpha_{(1)}} 
                 + \frac{\omega_{(3)m_2}}{\mu\alpha_{(3)}} \right)
		 \tilde{N}^{(1,3)}_{n_1,m_1}
		 \tilde{N}^{(1,3)}_{-n_2,m_2}
		 \tilde{N}^{(1,3)}_{n_3,m_3}
		 \tilde{N}^{(1,3)}_{-n_{123},-m_{123}} ,\\
\Pcm{N}_3&=&\left( \frac{\omega_{(1)n_3}}{\mu\alpha_{(1)}} 
                 + \frac{\omega_{(3)m_3}}{\mu\alpha_{(3)}} \right)
		 \tilde{N}^{(1,3)}_{n_1,m_1}
		 \tilde{N}^{(1,3)}_{n_2,m_2}
		 \tilde{N}^{(1,3)}_{-n_3,m_3}
		 \tilde{N}^{(1,3)}_{-n_{123},-m_{123}} ,\\
\Pcm{N}_4&=&\left( \frac{\omega_{(1)-n_{123}}}{\mu\alpha_{(1)}} 
                 + \frac{\omega_{(3)-m_{123}}}{\mu\alpha_{(3)}} \right)
		 \tilde{N}^{(1,3)}_{n_1,m_1}
		 \tilde{N}^{(1,3)}_{n_2,m_2}
		 \tilde{N}^{(1,3)}_{n_3,m_3}
		 \tilde{N}^{(1,3)}_{n_{123},-m_{123}}
,\end{eqnarray}
where the string frequencies are $\omega_{(r)m} = \sqrt{m^2+\mu^2\alpha_{(r)}^2}$.
The $\alpha_{(r)}$ are the fractions of outgoing light-cone
momentum carried by each string, and in
the present case these are $\alpha_{(1)}=1-s$, $\alpha_{(2)}=s$ and $\alpha_{(3)}=-1$.
The Neumann coefficients for the plane-wave geometry are given 
by \cite{0211198}, and we display them here for convenience ($m,n>0$):
\begin{eqnarray}
\tilde{N}^{(r,s)}_{0,n}&=&\tilde{N}^{(r,s)}_{0,-n}=\frac1{\sqrt{2}}\bar{N}^{(r,s)}_{0,n} \nn\\
\tilde{N}^{(r,s)}_{\pm m,\pm n}&=&\frac12\big( \bar{N}^{(r,s)}_{m,n} 
                                           - \bar{N}^{(r,s)}_{-m,-n} \big) \nn\\
\tilde{N}^{(r,s)}_{\pm m,\mp n}&=&\frac12\big( \bar{N}^{(r,s)}_{m,n} 
                                           + \bar{N}^{(r,s)}_{-m,-n} \big) \nn\\
\bar{N}^{(r,s)}_{0,n}&=&\frac{1}{2\pi}(-1)^{s(n+1)}s_{(s)n}
                        \sqrt{\frac{|\alpha_{(s)}|}{\alpha_{(r)}\omega_{(s)n}
                                 (\omega_{(s)n}+\mu\alpha_{(s)})}} \nn\\
\bar{N}^{(r,s)}_{\pm m,\pm n}&=&\pm\frac{1}{2\pi}\frac{(-1)^{r(m+1)+s(n+1)}s_{(r)m}s_{(s)n}}
                                  {\alpha_{(s)}\omega_{(r)m}+\alpha_{(r)}\omega_{(s)n}} \nn\\
&&\times\sqrt{\frac{|\alpha_{(r)}\alpha_{(s)}|
                    (\omega_{(r)m}\pm\mu\alpha_{(r)})(\omega_{(s)n}\pm\mu\alpha_{(s)})  }
                   {\omega_{(r)m}\omega_{(s)n}}
             } \\
s_{(1)m}&=&s_{(2)m}=1 \qquad s_{(3)m}=2\sin\big(\pi m \frac{\alpha_{(1)}}{\alpha_{(3)}}\big)
\end{eqnarray}
Expanding to leading order in $1/\mu$, the Neumann matrices become (for the cases we need)
\begin{equation}
\label{Neuexp}
\tilde{N}^{(r,3)}_{m,n}=\frac{(-1)^{r(m+1)+n}\sin(\pi ns)}{2\pi\sqrt{\alpha_{(r)}}
                          \big( \frac{m}{\alpha_{(r)}}-n \big)}
                  +\Pcm{O}\big(\frac1{\mu^2}\big)
.\end{equation}
Substituting into our expression \eqref{SFTff} for the `4-4' string amplitude, we obtain
\begin{eqnarray}
\label{SFToff}
\langle\textup{4,4}|H_3\rangle&=&\frac12\frac{s}{1-s}
\frac{(D_1)^2+(D_2)^2+(D_3)^2+(D_{123})^2}{D_1D_2D_3D_{123}} \nn\\
&&\times \sin(\frac{\pi s n_1}{r_0})
\sin(\frac{\pi s n_2}{r_0})
\sin(\frac{\pi s n_3}{r_0})
\sin(\frac{\pi s n_{123}}{r_0}) \nn\\
\textup{where } D_a &=& \frac{n_a}{r_0}-\frac{m_a}{r_0-s}
.\end{eqnarray}
Apart from normalisation, this is in complete agreement with the 
`4-4' matrix element \eqref{HTff} of the string Hamiltonian $\tilde{H}$ 
calculated on the SYM side, 
A calculation similar to the above leads to
\begin{eqnarray}
\label{SFTfto}
\langle\textup{4,3,1}|H_3\rangle&=&\frac{s(1-s)}2\bigg\{ 
            \left( \frac{\omega_{(3)n_1}}{\mu\alpha_{(3)}} 
                 + \frac{\omega_{(2)0}}{\mu\alpha_{(2)}} \right)
		 \tilde{N}^{(3,2)}_{-n_1,0}
		 \tilde{N}^{(3,1)}_{n_2,m_1}
		 \tilde{N}^{(3,1)}_{n_3,m_2}
		 \tilde{N}^{(3,1)}_{-n_{123},-m_{12}} \\
         &&{}+ \left( \frac{\omega_{(3)n_2}}{\mu\alpha_{(3)}} 
                 + \frac{\omega_{(1)m_1}}{\mu\alpha_{(1)}} \right)
		 \tilde{N}^{(3,2)}_{n_1,0}
		 \tilde{N}^{(3,1)}_{-n_2,m_1}
		 \tilde{N}^{(3,1)}_{n_3,m_2}
		 \tilde{N}^{(3,1)}_{-n_{123},-m_{12}} \\
         &&{}+ \left( \frac{\omega_{(3)n_3}}{\mu\alpha_{(3)}}
                 + \frac{\omega_{(1)m_2}}{\mu\alpha_{(1)}} \right)
		 \tilde{N}^{(3,2)}_{n_1,0}
		 \tilde{N}^{(3,1)}_{n_2,m_1}
		 \tilde{N}^{(3,1)}_{-n_3,m_2}
		 \tilde{N}^{(3,1)}_{-n_{123},-m_{12}} \\
         &&{}+ \left( \frac{\omega_{(3)-n_{123}}}{\mu\alpha_{(3)}}
                 + \frac{\omega_{(1)-m_{12}}}{\mu\alpha_{(1)}} \right)
		 \tilde{N}^{(3,2)}_{n_1,0}
		 \tilde{N}^{(3,1)}_{n_2,m_1}
		 \tilde{N}^{(3,1)}_{n_3,m_2}
		 \tilde{N}^{(3,1)}_{n_{123},-m_{12}} \bigg\}
,\end{eqnarray}
which upon substitution of the expanded Neumann matrices \eqref{Neuexp} leads to
\begin{eqnarray}
\label{SFTofto}
\langle\textup{4,3,1}|H_3\rangle&=&\frac12\sqrt{\frac{s}{(1-s)}}
\frac{(D_{123}^{-12})^2+(D_{1}^{-1})^2+(D_{2}^{-2})^2+(D_3)^2}
 {D_{123}^{-12}D_{1}^{-1}D_{2}^{-2}D_3} \nn\\
&&\times \sin(\frac{\pi s n_1}{r_0})
\sin(\frac{\pi s n_2}{r_0})
\sin(\frac{\pi s n_3}{r_0})
\sin(\frac{\pi s n_{123}}{r_0}) \nn\\
\textup{where } && D_a^b = \frac{n_a}{r_0} + \frac{m_b}{r_0-s}
.\end{eqnarray}
Again, this is in perfect agreement with the gauge-theory result \eqref{HTfto}.
Finally, we calculate the `4-22' interaction
\begin{eqnarray}
\label{SFTftt}
\langle\textup{4,2,2}|H_3\rangle&=&\frac{s(1-s)}2\bigg\{  
   \left( \frac{\omega_{(3)n_1}}{\mu\alpha_{(3)}} 
                 + \frac{\omega_{(1)m_1}}{\mu\alpha_{(1)}} \right)
		 \tilde{N}^{(3,2)}_{-n_1,m_1}
		 \tilde{N}^{(3,1)}_{n_2,-m_1}
		 \tilde{N}^{(3,2)}_{n_3,m_2}
		 \tilde{N}^{(3,2)}_{-n_{123},-m_{2}} \\
         &&{}+\left( \frac{\omega_{(3)n_2}}{\mu\alpha_{(3)}} 
                 + \frac{\omega_{(1)-m_1}}{\mu\alpha_{(1)}} \right)
		 \tilde{N}^{(3,2)}_{n_1,m_1}
		 \tilde{N}^{(3,1)}_{-n_2,-m_1}
		 \tilde{N}^{(3,2)}_{n_3,m_2}
		 \tilde{N}^{(3,2)}_{-n_{123},-m_{2}} \\
         &&{}+\left( \frac{\omega_{(3)n_3}}{\mu\alpha_{(3)}} 
                 + \frac{\omega_{(2)-m_2}}{\mu\alpha_{(2)}} \right)
		 \tilde{N}^{(3,2)}_{n_1,m_1}
		 \tilde{N}^{(3,1)}_{n_2,-m_1}
		 \tilde{N}^{(3,2)}_{-n_3,m_2}
		 \tilde{N}^{(3,2)}_{-n_{123},-m_{2}} \\
         &&{}+\left( \frac{\omega_{(3)-n_{123}}}{\mu\alpha_{(3)}} 
                 + \frac{\omega_{(2)-m_2}}{\mu\alpha_{(2)}} \right)
		 \tilde{N}^{(3,2)}_{n_1,m_1}
		 \tilde{N}^{(3,1)}_{n_2,-m_1}
		 \tilde{N}^{(3,2)}_{n_3,m_2}
		 \tilde{N}^{(3,2)}_{n_{123},-m_{2}} \bigg\}
,\end{eqnarray}
and find
\begin{eqnarray}
\label{SFToftt}
\langle\textup{4,2,2}|H_3\rangle&=&\frac12\sqrt{\frac{1-s}{s}}
\frac{(D_{123}^{-1})^2+(D_{2}^{2})^2+(D_{3}^{-2})^2+(D_1^{-1})^2}
{D_{123}^{-1}D_{2}^{2}D_{3}^{-2}D_1^{-1}} \nn\\
&&\times \sin(\frac{\pi s n_1}{r_0})
\sin(\frac{\pi s n_2}{r_0})
\sin(\frac{\pi s n_3}{r_0})
\sin(\frac{\pi s n_{123}}{r_0}) \nn\\
\textup{where }&& D_a^1 = \frac{n_a}{r_0} + \frac{m_1}{r_0-s} \textup{ and }
                  D_a^2 = \frac{n_a}{r_0} + \frac{m_2}{s}
;\end{eqnarray}
agreement with \er{HTftt} obtains.

We see that the string 
Hamiltonian $\tilde{H}$ calculated on the Yang-Mills
side for four-impurity BMN states reproduces the light-cone string 
field vertex between four-excitation string states.

\section{Discussion}

We have used the dilatation operator on the gauge-theory side of 
the correspondence in the basis of four-impurity BMN operators
to derive an expression for the corresponding Hamiltonian on the
string side. Four-impurity operators, as in the two- and 
three-impurity cases, may be considered to
have distinct impurities, and this fact has been used to
simplify the calculations. Nevertheless, to obtain the gauge-theory
matrix elements and transform to the momentum-state string basis
requires many tedious pages of calculation. This is rewarded by
the precise agreement between these and the matrix elements
obtained directly from the string field theory vertex in the
plane-wave background. Our results are also in agreement
with the perturbative gauge theory analysis in \cite{0306234,0301250}.
One could follow the analysis presented in \cite{0402211} and calculate 
decay widths using our matrix elements of $\tilde{H}$, although
it seems that only special cases may be treated analytically.

Our analysis gives further evidence
of the correspondence of BMN operators to string oscillator states,
although we have not addressed the potential discrepancy recently
uncovered at three-loop order\cite{0307032,0404007}.

It would be interesting to understand in detail what the effect
of repeated impurities would be, and to extend our calculations 
explicitly to consider more than four. In this case, there will
be more possible contractions of the gauge-theory operators, since
there will no longer be a unique contraction of the impurity fields.
These calculations could be compared against the arbitrary 
impurity-number calculations of \cite{0306234,0301250}.


\section*{Acknowledgements}

This work was supported in part by an operating grant from the 
National Sciences and Engineering Research Council of Canada.
P.~M. wishes to acknowledge support from Simon Fraser University
in the form of a Graduate Fellowship.

\section*{Appendix I: Matrix Elements of $H_1$}

In this appendix we first calculate the action of $H_1$ on the
discrete basis states \eqref{dbasisfour}-\eqref{dbasisoneoneoneone}
and demonstrate the use of these results to find the 
matrix elements of $H_1$ in
the momentum basis \eqref{mbasisfour}-\eqref{mbasistwooneone}.
We find:

\begin{eqnarray}
H_+ \Pcm{O}_{p_ap_bp_cp_d}^{abcd}\prod_j\Pcm{O}_{J_j}
 &=& \frac{-\lambda'}{8\pi^2} \bigg[ \sum_{i=1}^{p_1-1}
          \big(
	       \Pcm{O}^{abcd}_{p_a-i-1,p_b,p_c,p_d+1}
	      +\Pcm{O}^{abcd}_{p_a-i-1,p_b+1,p_c,p_d}
	     -2\Pcm{O}^{abcd}_{p_a-i,p_b,p_c,p_d}
          \big)\Pcm{O}_i  \nn\\
  &+& \sum_{i=0}^{p_1-1} \bigg(
         \big(
	       \Pcm{O}^{bcd}_{p_b,p_c+1,p_a-i-1}
	      -\Pcm{O}^{bcd}_{p_b,p_c,p_a-i}
          \big)\Pcm{O}^a_{p_d+i} \nn\\
  &&      +\big(
	       \Pcm{O}^{ad}_{i,p_d+1}
	      -\Pcm{O}^{ad}_{i+1,p_d}
          \big)\Pcm{O}^{bc}_{p_b,p_c+p_a-i-1} \nn\\
  &&      +\big(
	       \Pcm{O}^{bc}_{p_b+1,p_a-i-1}
	      -\Pcm{O}^{bc}_{p_b,p_a-i}
          \big)\Pcm{O}^{ad}_{p_c+i,p_d} \nn\\
  &&      +\big(
	       \Pcm{O}^{acd}_{i,p_c+1,p_d}
	      -\Pcm{O}^{acd}_{i+1,p_c,p_d}
          \big)\Pcm{O}^{b}_{p_b+p_a-i-1} \bigg) \bigg] \prod_j\Pcm{O}_{J_j} \nn\\
    & + & (\text{three other cyclic permutations of }abcd) ,
\end{eqnarray}
\begin{eqnarray}
H_- \Pcm{O}_{p_ap_bp_cp_d}^{abcd}\prod_j\Pcm{O}_{J_j}
  &=& \frac{-\lambda'}{8\pi^2} \bigg[ \sum_{i=1}^{l} J_i \big(
	       \Pcm{O}^{abcd}_{p_a+J_i-1,p_b,p_c,p_d+1}
	      -\Pcm{O}^{abcd}_{p_a+J_i,p_b,p_c,p_d} \nn\\
    &&	      +\Pcm{O}^{abcd}_{p_a+1,p_b,p_c,p_d+J_i-1}
      	      -\Pcm{O}^{abcd}_{p_a,p_b,p_c,p_d+J_i} 
	      \big) \prod_{j\ne i}\Pcm{O}_{J_j} \bigg]\nn\\
    & +& (\text{three other cyclic permutations of }abcd)  .
\end{eqnarray}
\begin{eqnarray}
H_+ \Pcm{O}_{p_ap_bp_c}^{abc}\Pcm{O}^d_{p_d}\prod_j\Pcm{O}_{J_j}
 &=& \frac{-\lambda'}{8\pi^2} \bigg[ \sum_{i=1}^{p_a-1}
          \big(
	       \Pcm{O}^{abc}_{p_a-i-1,p_b,p_c+1}
	      +\Pcm{O}^{abc}_{p_a-i-1,p_b+1,p_c}
             -2\Pcm{O}^{abc}_{p_a-i,p_b,p_c}
          \big)\Pcm{O}^d_{p_d}\Pcm{O}_i  \nn\\
 &+&  \sum_{i=0}^{p_a-1} \bigg(
  	 \big(\Pcm{O}^{cb}_{p_a-i-1,p_b+1}
              -\Pcm{O}^{cb}_{p_a-i,p_b}
          \big)\Pcm{O}^{a}_{p_c+i}\Pcm{O}^{d}_{p_d} \nn\\
  &&	 +\big(\Pcm{O}^{ca}_{p_c+1,p_a-i-1}
              -\Pcm{O}^{ca}_{p_c,p_a-i}
          \big)\Pcm{O}^{b}_{p_b+i}\Pcm{O}^{d}_{p_d}
	      \bigg) \bigg]\prod_j\Pcm{O}_{J_j} \nn\\
     &+& (\text{two other cyclic permutations of }abc) ,
\end{eqnarray}
\begin{eqnarray}
H_- \Pcm{O}_{p_ap_bp_c}^{abc}\Pcm{O}^d_{p_d}\prod_j\Pcm{O}_{J_j}
 &=& \frac{-\lambda'}{8\pi^2} \bigg[\sum_{i=1}^{l}  J_i \big(
	       \Pcm{O}^{abc}_{J_i+p_a-1,p_b,p_c+1}
	      -\Pcm{O}^{abc}_{J_i+p_a,p_b,p_c} \nn\\
	 &&   +\Pcm{O}^{abc}_{p_a+1,p_b,p_c+J_i-1}
	      -\Pcm{O}^{abc}_{p_a,p_b,p_c+J_i}
              \big) \Pcm{O}^d_{p_d} \nn\\
 &+&  \sum_{i=0}^{p_a-1} \bigg(
	       \Pcm{O}^{dbca}_{p_a-i-1,p_b,p_c,p_d+i+1}
	      -\Pcm{O}^{dbca}_{p_a-i,p_b,p_c,p_d+i}   \nn\\
  &&	      +\Pcm{O}^{dbca}_{p_d+i+1,p_b,p_c,p_a-i-1}
	      -\Pcm{O}^{dbca}_{p_d+i,p_b,p_c,p_a-i}  
	      \bigg) \Pcm{O}_{J_i} \nn\\
  &+& \sum_{i=0}^{p_d-1}
          \big(
	       \Pcm{O}^{adbc}_{p_d-i-1,p_a+i,p_b,p_c+1}
	      -\Pcm{O}^{adbc}_{p_d-i,p_a+i,p_b,p_c}  \nn\\
	  &&  +\Pcm{O}^{abcd}_{p_a+1,p_b,p_c+i,p_d-i-1}
	      -\Pcm{O}^{abcd}_{p_a,p_b,p_c+i,p_d-i}
          \big) \Pcm{O}_{J_i} 
  \bigg]\prod_{j\ne i}\Pcm{O}_{J_j}  \nn\\
    &+& (\text{two other cyclic permutations of }abc) .
\end{eqnarray}
\begin{eqnarray}
H_+ \Pcm{O}_{p_ap_b}^{ab}\Pcm{O}^{cd}_{p_cp_d}\prod_j\Pcm{O}_{J_j}
  &=& \frac{-\lambda'}{8\pi^2} \sum_{i=1}^{p_a-1}
          2\big(
	       \Pcm{O}^{ab}_{p_a-i-1,p_b+1}
	      -\Pcm{O}^{ab}_{p_a-i,p_b}
          \big)\Pcm{O}^{cd}_{p_cp_d}\Pcm{O}_i 
       \prod_j\Pcm{O}_{J_j} \nn\\
  &+& \left( \begin{array}{c} a \leftrightarrow c \\ b \leftrightarrow d \end{array} 
    \right) \text{, then } + (a \leftrightarrow b) \text{, then } + (c \leftrightarrow d),
\end{eqnarray}
\begin{eqnarray}
H_- \Pcm{O}_{p_ap_b}^{ab}\Pcm{O}^{cd}_{p_cp_d}\prod_j\Pcm{O}_{J_j}
  &=& \frac{-\lambda'}{8\pi^2} \bigg[\sum_{i=1}^{l} J_i \big(
	       \Pcm{O}^{ab}_{J_i+p_a-1,p_b+1}
	      -\Pcm{O}^{ab}_{J_i+p_a,p_b}  \nn\\
	&&    +\Pcm{O}^{ab}_{p_a+1,p_b+J_i-1}
	      -\Pcm{O}^{ab}_{p_a,p_b+J_i}
              \big) \Pcm{O}^{cd}_{p_cp_d} \nn\\
  &+& \sum_{i=0}^{p_a-1} \big(
	       \Pcm{O}^{cdba}_{p_c+1,p_d+i,p_b,p_a-i-1}
	      -\Pcm{O}^{cdba}_{p_c,p_d+i,p_b,p_a-i} \nn\\
	&&    +\Pcm{O}^{cbad}_{p_a-i-1,p_b,p_c+i,p_d+1}
	      -\Pcm{O}^{cbad}_{p_a-i,p_b,p_c+i,p_d}
	      \big)\Pcm{O}_{J_i} \bigg] \prod_{j\ne i}\Pcm{O}_{J_j}  \nn\\
    &+& \left( \begin{array}{c} a \leftrightarrow c \\ b \leftrightarrow d \end{array}
    \right) \text{, then } + (a \leftrightarrow b) \text{, then } + (c \leftrightarrow d).
\end{eqnarray}
\begin{eqnarray}
H_+ \Pcm{O}_{p_ap_b}^{ab}\Pcm{O}_{p_c}^{c}\Pcm{O}_{p_d}^{d}\prod_j\Pcm{O}_{J_j}
  &=& \frac{-\lambda'}{8\pi^2} \bigg[
   -\frac12\sum_{i=1}^{p_a-1} 
         \big(
	       \Pcm{O}^{ab}_{p_a-i,p_b}
	      -\Pcm{O}^{ab}_{p_a-i-1,p_b+1} \nn\\
	&&    -\Pcm{O}^{ab}_{p_a-1,p_b-i+1}
	      +\Pcm{O}^{ab}_{p_a,p_b-i}
          \big)\Pcm{O}^c_{p_c}\Pcm{O}^d_{p_d}\Pcm{O}_i \bigg] \prod_j\Pcm{O}_{J_j} \nn\\ 
	&+&  (c \leftrightarrow d) \text{, then } + (a \leftrightarrow b) ,
\end{eqnarray}
\begin{eqnarray}
H_- \Pcm{O}_{p_ap_b}^{ab}\Pcm{O}_{p_c}^{c}\Pcm{O}_{p_d}^{d}\prod_j\Pcm{O}_{J_j}
  &=& \frac{-\lambda'}{8\pi^2} \bigg[-\frac12\sum_{i=1}^{l}  J_i \big(
	       \Pcm{O}^{ab}_{J_i+p_a,p_b}
	      -\Pcm{O}^{ab}_{J_i+p_a-1,p_b+1}  \nn\\
	&&    +\Pcm{O}^{ab}_{p_a,p_b+J_i}
	      -\Pcm{O}^{ab}_{p_a+1,p_b+J_i-1}
              \big) \Pcm{O}^{c}_{p_c}\Pcm{O}^{d}_{p_d} \nn\\
  &-&\sum_{i=0}^{p_c-1} \bigg(
          \big(
	       \Pcm{O}^{abc}_{p_a,p_c-i,p_b+i}
	      -\Pcm{O}^{abc}_{p_a+1,p_c-i-1,p_b+i} \nn\\
	&&    +\Pcm{O}^{abc}_{p_a,p_b+i,p_c-i}
	      -\Pcm{O}^{abc}_{p_a+1,p_b+i,p_c-i-1}
          \big)\Pcm{O}^d_{p_d} \nn\\
  && + \frac12  \big(
	       \Pcm{O}^{dc}_{p_c-i,p_d+i}
	      -\Pcm{O}^{dc}_{p_c-i-1,p_d+i+1} \nn\\
	&&    +\Pcm{O}^{dc}_{p_d+i,p_c-i}
	      -\Pcm{O}^{dc}_{p_d+i+1,p_c-i-1} 
          \big)\Pcm{O}^{ab}_{p_ap_b} \bigg)\Pcm{O}_{J_i} \nn\\
   &-&\sum_{i=0}^{p_a-1}
          \big(
	       \Pcm{O}^{cba}_{p_a-i,p_b,p_c+i}
	      -\Pcm{O}^{cba}_{p_a-i-1,p_b,p_c+i+1} \nn\\
	&&    +\Pcm{O}^{cba}_{p_a+i,p_b,p_c-i}
	      -\Pcm{O}^{cba}_{p_a+i+1,p_b,p_c-i-1}
          \big)\Pcm{O}^d_{p_d}\Pcm{O}_{J_i} 
	  \bigg] \prod_{j\ne i}\Pcm{O}_{J_j}  \nn\\
	&+&  (c \leftrightarrow d) \text{, then } + (a \leftrightarrow b) .
\end{eqnarray}
\begin{equation}
H_+ \Pcm{O}_{p_a}^{a}\Pcm{O}_{p_b}^{b}\Pcm{O}_{p_c}^{c}\Pcm{O}_{p_d}^{d}\prod_j\Pcm{O}_{J_j} = 0.
\end{equation}
\begin{eqnarray}
H_- \Pcm{O}_{p_a}^{a}\Pcm{O}_{p_b}^{b}\Pcm{O}_{p_c}^{c}\Pcm{O}_{p_d}^{d}\prod_j\Pcm{O}_{J_j}
  &=& \frac{-\lambda'}{8\pi^2} \bigg[\sum_{i=0}^{p_b-1} 
          \big(
	       \Pcm{O}^{ab}_{p_b-i,p_a+i}
	      -\Pcm{O}^{ab}_{p_b-i-1,p_a+i+1} \nn\\
	&&    +\Pcm{O}^{ab}_{p_a+i,p_b-i}
	      -\Pcm{O}^{ab}_{p_a+i+1,p_b-i-1}
          \big)\Pcm{O}^c_{p_c}\Pcm{O}^d_{p_d} \bigg] \prod_j\Pcm{O}_{J_j}\nn\\
  &+& (\text{11 other permutations of } abcd) ,
\end{eqnarray}

As with the case of $H_0$, we now write the continuum forms. 
In order to save space, we will not
write out all the permutations, instead using the convention that in the 
following the permutations must first be carried out,
and then $x_d$ set to $r_0-x_{abc}$ and $\del_d$ set to zero. 
Since they are unaffected by $H_+$, we suppress factors of $\prod\Pcm{O}(r_j)$
in the expressions for $H_+$.
We find:

`4'
\begin{eqnarray}
\label{hpfour}
 && H_+\Pcm{O}^{abcd}(x_a,x_b,x_c,x_d) = \frac{-\lambda'}{8\pi^2} \nn\\
 && \int_0^{x_a}\td y \big[ 
  (\del_d-2\del_a+\del_b)\Pcm{O}^{abcd}(x_a-y,x_b,x_c,x_d)\Pcm{O}(y) \nn\\
 && +(\del_c-\del_a)\Pcm{O}^{bcd}(x_b,x_c,x_a-y)\Pcm{O}^{a}(x_d+y) \nn\\
 && +(\del_d-\del_a)\Pcm{O}^{ad}(x_a-y,x_d)\Pcm{O}^{bc}(x_b+x_c+y) \nn\\
 && +(\del_b-\del_a)\Pcm{O}^{bc}(x_b,x_a-y)\Pcm{O}^{ad}(x_c+y,x_d) \nn\\
 && +(\del_c-\del_a)\Pcm{O}^{acd}(x_a-y,x_c,x_d)\Pcm{O}^{b}(x_b+y) \big] \nn\\
 &&   + (\textup{3 other cyclic permutations of }abcd)
\end{eqnarray}
\begin{eqnarray}
&&H_-\Pcm{O}^{abcd}(x_a,x_b,x_c,x_d) \prod_j\Pcm{O}(r_j) = \frac{-\lambda'}{8\pi^2}  \nn\\
&& \sum_{i=1}^{l}r_i
    (\del_d-\del_a)(\Pcm{O}^{abcd}(x_a+r_i,x_b,x_c,x_d)-\Pcm{O}^{abcd}(x_a,x_b,x_c,x_d+r_i))
    \prod_{j\ne i}\Pcm{O}(r_j) \nn\\
 &&   + (\textup{3 other cyclic permutations of }abcd)
\end{eqnarray}

`31'
\begin{eqnarray}
&&H_+\Pcm{O}^{abc}(x_a,x_b,x_c)\Pcm{O}^{d}(x_d) = \frac{-\lambda'}{8\pi^2}  \nn\\
  && \int_0^{x_a} \td y \big[
   (\del_c-2\del_a+\del_b)\Pcm{O}^{abc}(x_a-y,x_b,x_c)\Pcm{O}^d(x_d)\Pcm{O}(y) \nn\\
  && +(\del_b-\del_a)\Pcm{O}^{cb}(x_a-y,x_b)\Pcm{O}^{a}(x_c+y)\Pcm{O}^d(x_d)
   +(\del_c-\del_a)\Pcm{O}^{ca}(x_c,x_a-y)\Pcm{O}^{b}(x_b+y)\Pcm{O}^d(x_d) \big] \nn\\
  && +(\textup{2 other cyclic permutations of }abc)
\end{eqnarray}
\begin{eqnarray}
&&H_-\Pcm{O}^{abc}(x_a,x_b,x_c) \Pcm{O}^d(x_d) \prod_j\Pcm{O}(r_j) = \frac{-\lambda'}{8\pi^2}  \nn\\
&& \sum_{i=1}^{l}r_i
    (\del_c-\del_a)(\Pcm{O}^{abc}(x_a+r_i,x_b,x_c)-\Pcm{O}^{abc}(x_a,x_b,x_c+r_i))\Pcm{O}^d(x_d)
    \prod_{j\ne i}\Pcm{O}(r_j) \nn\\
  &+& \int_0^{x_a} \td y \big[
   (\del_d-\del_a)\Pcm{O}^{dbca}(x_d+y,x_b,x_c,x_a-y) 
     + (\del_d-\del_a)\Pcm{O}^{dbca}(x_d-y,x_b,x_c,x_a+y)\big] \prod_j\Pcm{O}(r_j)\nn\\
 &+&\int_0^{x_d} \td y \big[
   (\del_c-\del_d)\Pcm{O}^{adbc}(x_d-y,x_a+y,x_b,x_c)
   +(\del_a-\del_d)\Pcm{O}^{abcd}(x_a,x_b,x_c+y,x_d-y)\big]\prod_j\Pcm{O}(r_j) \nn\\
  && +(\textup{2 other cyclic permutations of }abc)
\end{eqnarray}

`22'
\begin{eqnarray}
&&H_+\Pcm{O}^{ab}(x_a,x_b)\Pcm{O}^{cd}(x_c,x_d) = \frac{-\lambda'}{8\pi^2}  \nn\\
  && \int_0^{x_a} \td y 
   2(\del_c-\del_a)\Pcm{O}^{ab}(x_a-y,x_b)\Pcm{O}^{cd}(x_c,x_d)\Pcm{O}(y) \nn\\
 && +(a \leftrightarrow c , b \leftrightarrow d) 
     , \textup{ then}  + (a \leftrightarrow b) 
     , \textup{ then} + (c \leftrightarrow d)
\end{eqnarray}
\begin{eqnarray}
&&H_-\Pcm{O}^{ab}(x_a,x_b) \Pcm{O}^{cd}(x_c,x_d) \prod_j\Pcm{O}(r_j) = \frac{-\lambda'}{8\pi^2} \nn\\
&& \sum_{i=1}^{l}r_i
    (\del_a-\del_b)(\Pcm{O}^{ab}(x_a,x_b+r_i)-\Pcm{O}^{ab}(x_a+r_i,x_b))\Pcm{O}^{cd}(x_c,x_d)
    \prod_{j\ne i}\Pcm{O}(r_j) \nn\\
  &+& \int_0^{x_a} \td y 
    \big[ (\del_c-\del_a)\Pcm{O}^{cdba}(x_c,x_d,x_b,x_a) \nn\\
  && +(\del_d-\del_a)\Pcm{O}^{cbad}(x_a-y,x_b,x_c+y,x_d) \big] \prod_j\Pcm{O}(r_j)\nn\\
 && +(a \leftrightarrow c , b \leftrightarrow d) 
     , \textup{ then}  + (a \leftrightarrow b) 
     , \textup{ then} + (c \leftrightarrow d)
\end{eqnarray}

`211'
\begin{eqnarray}
&&H_+\Pcm{O}^{ab}(x_a,x_b)\Pcm{O}^{c}(x_c)\Pcm{O}^{d}_{x_d} = \frac{-\lambda'}{8\pi^2}  \nn\\
  && \frac12\int_0^{x_a} \td y 
    (\del_b-\del_a) (\Pcm{O}^{ab}(x_a-y,x_b)
    + \Pcm{O}^{ab}(x_a,x_b-y) ) \Pcm{O}^c(x_c)\Pcm{O}^d(x_d)\Pcm{O}(y) \nn\\
&& +(c \leftrightarrow d) , \textup{ then} + (a \leftrightarrow b)
\end{eqnarray}
\begin{eqnarray}
&&H_-\Pcm{O}^{ab}(x_a,x_b) \Pcm{O}^{c}(x_c) \Pcm{O}^{d}(x_d) \prod_j\Pcm{O}(r_j) = \frac{-\lambda'}{8\pi^2} \nn\\
&& \frac12 \sum_{i=1}^{l}r_i
    (\del_a-\del_b)(\Pcm{O}^{ab}(x_a,x_b+r_i)-\Pcm{O}^{ab}(x_a+r_i,x_b))\Pcm{O}^{c}(x_c)\Pcm{O}^{d}(x_d)
    \prod_{j\ne i}\Pcm{O}(r_j) \nn\\
  &+& \int_0^{x_c} \td y \big[
     (\del_a-\del_c)(   \Pcm{O}^{abc}(x_a,x_c-y,x_b+y)
                        + \Pcm{O}^{abc}(x_a,x_b+y,x_c-y)  ) \Pcm{O}^d(x_d) \nn\\
  &&+\frac12(\del_d-\del_c)(   \Pcm{O}^{dc}(x_c-y,x_d+y)
                               + \Pcm{O}^{dc}(x_d+y,x_d+y,x_c-y)  ) \Pcm{O}^{ab}(x_a,x_b) \big] \prod_j\Pcm{O}(r_j) \nn\\
  &+& \int_0^{x_a} \td y 
     (\del_a-\del_c)(   \Pcm{O}^{cba}(x_a+y,x_b,x_c-y)
                        - \Pcm{O}^{cba}(x_a-y,x_b,x_c+y)  ) \Pcm{O}^d(x_d) \prod_j\Pcm{O}(r_j)\nn\\
&& +(c \leftrightarrow d) , \textup{ then} + (a \leftrightarrow b)
\end{eqnarray}

`1111'
\begin{equation}
H_+\Pcm{O}^{a}(x_a) \Pcm{O}^{b}(x_b) \Pcm{O}^{c}(x_c) \Pcm{O}^{d}(x_d) \prod_j\Pcm{O}(r_j) =0
\end{equation}
\begin{eqnarray}
\label{hmoooo}
&&H_-\Pcm{O}^{a}(x_a)\Pcm{O}^{b}(x_b)\Pcm{O}^{c}(x_c)\Pcm{O}^{d}_{x_d} = \frac{-\lambda'}{8\pi^2}  \nn\\
  &&\int_0^{x_b} \td y 
     (\del_b-\del_a)(  \Pcm{O}^{ab}(x_b-y,x_a+y)
                       + \Pcm{O}^{ab}(x_a+y,x_b-y)  ) \Pcm{O}^d(x_d) \Pcm{O}^c(x_c) \nn\\
  && +(\textup{11 other permutations of abcd})
\end{eqnarray}

We point out that the elements not involving four-impurity states are 
very similar to those calculated in the study of three-impurity 
states in \cite{0402211},
the only difference being that there appears an additional 
single-impurity state which, like the zero-impurity states, is 
unaffected by $H_1$.
As discussed in the main text, we need only consider elements involving `4' states.
We now restrict our attention to the novel `4-4', `4-31' and `4-22' matrix 
elements, and express these in the momentum-state basis. 

The `4-4' component of $H_+$ may be obtained from the four permutations of the 
first line of \er{hpfour}.
This may be written
\begin{eqnarray}
H_+\big|_\textup{`4-4'} |n_1,n_2,n_3;r_0\rangle^{1234}&=& 
   \frac{-\lambda'}{8\pi^2} \frac{-2\pi i}{\sqrt{(r_0-s)^3sr_0^3}}
    \sum_{n_1,n_2,n_3}
      \int_{x_{123}<r_0}\td^3x \nn \\ \bigg\{
      &-&(2n_1+n_2+n_3)\int_0^{x_1}\td s \frac1{r_0-s} 
        e^{\frac{2\pi i}{r_0-s} m_{123}s} \nn\\
      &+&(n_1-n_2) \int_0^{x_2}\td s  \frac1{r_0-s}
        e^{\frac{2\pi i}{r_0-s} m_{23}s} \nn\\
      &+&(n_2-n_3) \int_0^{x_3}\td s \frac1{r_0-s}
        e^{\frac{2\pi i}{r_0-s} m_{3}s} \nn\\
      &+&(n_1+n_2+2n_3) \int_0^{r_0-x_{123}}\td s \frac1{r_0-s} \bigg\}\nn\\
     && e^{\frac{2\pi i}{r_0}\big(D_1x_1+D_2x_{12}+D_3x_{123}\big)} \nn\\
     && |m_1,m_2,m_3;r_0-s\rangle^{1234}|s\rangle
,\end{eqnarray}
where $D_a=\frac{n_a}{r_0}-\frac{m_a}{r_0-s}$.
Here, we have not yet added states where the impurities $2,3,4$ are permuted along 
with the three momenta. We can express this component of $H_+$ in terms of the 
basis state \eqref{mbasisfour} by adding these permutations. In this case,
this involves adding five more terms, in which the impurities $2,3,4$ are permuted,
along with the momenta $n_1,n_2,n_3$ and also the momenta $m_1,m_2,m_3$.
Performing the $x$-integrations and adding the above permutations, 
our final result for the `4-4' matrix element of $H_+$ is given by
\begin{eqnarray}
 \langle s | \langle m_1,m_2,m_3;r_0 &-&s| H_+ | n_1,n_2,n_3;r_0\rangle \nn\\
   & =&   \frac{\lambda'}{2\pi^4} \frac{1}{\sqrt{(r_0-s)^3sr_0^3}}
        \sin\big(\frac{\pi s n_1}{r_0}\big)
        \sin\big(\frac{\pi s n_2}{r_0}\big)
        \sin\big(\frac{\pi s n_3}{r_0}\big)
        \sin\big(\frac{\pi s n_{123}}{r_0}\big) \nn\\
&\times& \frac{m_1D_1+m_2D_2+m_3D_3+m_{123}D_{123}}
	      {(r_0-y)D_1D_2D_3D_{123}} 
.\end{eqnarray}

The other components of $H_+$ and $H_-$ are calculated in similar fashion
using the continuum forms \eqref{hpfour}-\eqref{hmoooo}, and the results are given in 
the main text.

\section*{Appendix I\!I: Matrix Elements of $\Sigma$}

Here we consider matrix elements of $\Sigma$ in the momentum-state 
basis; these are found by first calculating simple two-point 
functions in the original discrete basis and then transforming.
For the present calculation, we need to find the matrix elements 
of $\Sigma$ corresponding to those calculated in Appendix I for $H_+$ and $H_-$.
Since $\Sigma$ is Hermitian in the momentum-state basis, we do not need
to distinguish between trace-number increasing and decreasing parts
(which other authors have labeled $\Sigma_+$ and $\Sigma_-$) since these
are simply related by conjugation.

\setlength\arraycolsep{0pt}
The `4-4' component of $\Sigma$ is found by considering the 
correlator
\begin{equation}
\langle 
\bar{\Pcm{O}}^{1234}_{q_1,q_2,q_3,q_4-k} \bar{\Pcm{O}}_k \Pcm{O}^{1234}_{p_1,p_2,p_3,p_4} \rangle
\end{equation}
which to genus-one order is given by
\begin{equation}
N^{p_{1234}+3}k(p_4-k-1)\delta_{p_1,q_1}\delta_{p_2,q_2}\delta_{p_3,q_3}\delta_{p_4,q_4}
.\end{equation}
Taking the continuum limit, we replace $p$ and $q$ with $x$ and $y$; then
transforming to the momentum-state basis requires a tedious procedure of 
integration similar to that used to calculate the elements of $H_+$ and $H_-$.

The result is
\begin{eqnarray}
  \langle s |
 &&\langle m_1,m_2,m_3;r_0 - s| \Sigma | n_1,n_2,n_3;r_0\rangle \nn\\
    &&= \frac{1}{\pi^4\sqrt{(r_0-s)^3sr_0^3}}
      \sin\big(\frac{\pi s n_1}{r_0}\big)
      \sin\big(\frac{\pi s n_2}{r_0}\big)
      \sin\big(\frac{\pi s n_3}{r_0}\big)
      \sin\big(\frac{\pi s n_{123}}{r_0}\big) \nn\\
&&\times \frac{1}{D_1D_2D_3D_{123}} \nn\\
\textup{where } && D_a = \frac{n_a}{r_0} - \frac{m_a}{r_0-s}
.\end{eqnarray}
Using the same method, the `4-31' element may be found from the correlator
\begin{equation}
\langle \bar{\Pcm{O}}^{123}_{q_1,q_2,q_3} \bar{\Pcm{O}}^4_{q_4} \Pcm{O}^{1234}_{p_1,p_2,p_3,p_4} \rangle
=N^{p_{1234}+3}(\textup{min}(p_3,q_3,p_4,q_4)+1)\delta_{p_1,q_1}\delta_{p_2,q_2}\delta_{p_3+p_4,q_3+q_4}
,\end{equation}
and the `4-22' element from the correlator
\begin{equation}
\langle \bar{\Pcm{O}}^{12}_{q_1,q_2} \bar{\Pcm{O}}^{34}_{q_3,q_4} \Pcm{O}^{1234}_{p_1,p_2,p_3,p_4} \rangle
=N^{p_{1234}+3}(\textup{min}(p_2,q_2,p_4,q_4)+1)\delta_{p_1,q_1}\delta_{p_3,q_3}\delta_{p_2+p_4,q_2+q_4}
.\end{equation}
The results are given in the main text.
\setlength\arraycolsep{2pt}

\end{document}